\documentclass[12pt,fleqn]{article}
\usepackage{amsmath}
\usepackage{cite}
\usepackage{amsfonts}
\newlength{\dinwidth}
\newlength{\dinmargin}
\begin{document}
\begin{flushright}
DFF 298/1/98
\end{flushright}
\vspace{1cm}

\begin{center}

{\Large{\textbf{Energy Scale and Coherence Effects in Small-\boldmath{$x$}
 Equations}}}

\vspace{1cm}
\normalsize{Marcello Ciafaloni} 
 \vspace{1cm}

\textit{Dipartimento di Fisica dell'Universit\`a, Firenze\\
and INFN, Sezione di Firenze, Italy} 

\end{center}
\vspace{1cm}

\begin{abstract}
I consider the next-to-leading high-energy cluster expansion of large-{\bf k} double jet
production in QCD, and I determine the corresponding one-loop quark and gluon impact 
factors for a self-consistent energy scale. The result shows that coherent
angular ordering of emitted gluons holds for hard emission also, and singles out a 
scale which in essentially the largest virtuality in the process. Both remarks are relevant for
the precise determination of the BFKL kernel at the next-to-leading level.
\end{abstract}
\vspace{1cm}
\begin{center}
\end{center}

\newpage

Various attempts to understand the small-$x$ HERA data on structure functions and final states \cite{1}
have stimulated the analysis of high-energy QCD beyond the leading-logarithmic approximation \cite{2}.

After several years of investigation, the high-energy vertices needed for the next-to-leading (NL)
approximation have been computed \cite{3,4,5,6,7,8}, and the irreducible part of the NL kernel of the BFKL 
equation \cite{2} has been written out explicitly \cite{9,10,11}. It turns out that the NL terms provide
(sizeable) corrections to the hard Pomeron intercept and to the singlet anomalous dimensions \cite{10}.

However, the definition of the NL kernel is not free of ambiguities which have 
prevented, so far, a
complete quantitative analysis of the NL results. The most important 
ambiguity is due to the dependence
of some NL features on the determination of the
 physical scale of the squared energy $s$, in the log $s$ dependence
of the cross sections.

In a hard process like DIS, the scale of $s$ is taken to be $Q^{2}$, 
the virtuality of the photon or
electroweak boson involved. Thus the structure functions are basically 
dependent on the Bjork\`{e}n scaling variable
$x = Q^{2}/s$, with scaling violations induced by $\alpha_{s}(Q^{2})$.
Similar considerations can be made for double DIS \cite{12} or quarkonium
production, where two hard scales are present.

On the other hand, the high-energy cluster expansion needed for the definition of the NL kernel,
has been mostly investigated in the case of parton-parton scattering, in which no physical hard scale
is present. In order to define such scale(s), we have to deal with a partial cross section, by fixing the
virtualities ${\bf k}_{1}, \mathbf{k}_{2}$ of the 
off-shell gluon Green's function by high-energy factorization (Fig. 1a).
Experimentally, this corresponds \cite{13} to the cross section associated with the production of
two jets with given transverse momenta $\mathbf{k}_{1}, \mathbf{k}_{2}$ and large relative rapidity, in the 
fragmentation regions of the incoming partons A, B.

We can then define the scale $s_{0} (\mathbf{k}_{1}, \mathbf{k}_{2})$ as the one occurring in the logarithmic behaviour
of the gluon Green's function $G(\mathbf{k}_{1}, \mathbf{k}_{2};s)$, whose loop expansion reads
\begin{equation}
G = 1+\left( K_{0}\log \frac {s}{s_{0}(\mathbf{k}_{1}, \mathbf{k}_{2})}+
\textrm{const.} \right)  
+\left(\frac{1}{2} K^{2}_{0} 
\left(\log \frac {s}{s_0}\right)^{2} 
+  K_{NL} \log \frac{s}{s_0} + \textrm{const.}\right) + \cdots
\label{(1)}
\end{equation}
where we have introduced the leading and next-to-leading operator kernels $K_{0}, K_{NL}$. It
becomes then clear that the determination of $K_{NL}$ on the basis of an explicit two-loop
calculation is dependent on the choice of $s_{0}$: a change in the latter is a NL effect which
induces a change in $K_{NL}$, for the expansion (\ref{(1)}) to be left unchanged.

More precisely, I will assume in the following a $\mathbf{k}$ - factorized form for the differential
cross section
\begin{equation}
\frac{d \sigma^{AB}} {d[\mathbf{k}_{1}] d[\mathbf{k}_{2}]} = \int \frac {d\omega} {2\pi i} \left(\frac {s} 
{s_0(\mathbf{k}_{1}, \mathbf{k}_{2})}\right)^{\omega} \frac{\pi}{\omega}h^{A}_{\omega}
(\mathbf{k}_1)G_{\omega}(\mathbf{k}_1, \mathbf{k}_2) \;
\; h^B_{\omega}(\mathbf{k}_2)
\label{fracd}      
\end{equation}
where $d[\mathbf{k}] = d^{2(1+\epsilon)}\mathbf{k}/ \pi^{1+\epsilon}$ in $D = 4+2\epsilon$ dimensions,
\begin{equation}
G_{\omega} = [1 - \frac{1}{\omega} (K_{0} + K_{NL})]^{-1}
\end{equation}
is the Mellin transform of the gluon Green's function \footnote{An alternative
formulation, not adopted here, could include an $\omega$-independent
NL factor in Eq.(3).} 
and $h^{A,B}$ are (process dependent)
impact factors, to be determined also, which may carry collinear singularities due to the 
initial (massless) partons A,B.

In this note I will present arguments for a choice of $s_{0}$ which is consistent with the collinear properties 
of the partonic process and I will give the corresponding one-loop determination of the
impact factors. These results can then be used to yield a precise determination of $K_{NL}$ at
two-loop level \cite{10}.

It turns out that the basic physical issue is the so-called ``coherent effect'' in the space-like jets, noticed
by the author long ago \cite{14}. This effect can be stated by saying that the leading log
squared matrix element for gluon emission holds in a dynamically restricted phase space, in
which angular ordering of the emitted gluons is required. 
The angular restriction determines, 
among other features, the energy scale also.

Let me start noting that the one-loop 
$Ag^{\ast}(\mathbf{k}_2)\rightarrow Ag^{\ast}(\mathbf{k}_2)$
impact factor involves the known \cite{5} particle-particle reggeon vertex 
$AA^{\prime} g^{\ast}(\mathbf{k}_2)$
at virtual level, and the $Ag^{\ast}(\mathbf{k}_2)\rightarrow A^{\prime} g(\mathbf{q})$ 
squared matrix element
(Fig. 1b) at real emission level. The latter quantity can be extracted from the exact squared matrix elements of Ellis 
and Sexton \cite{15}, and has been investigated by various authors at 
matrix element level \cite{3,8,16}
but has not yet been provided in explicit form in D dimensions, to the author's knowledge. I thus derive it 
directly for the quark case.

By using the notation of Fig. 1 and the Sudakov parametrization 
\begin{eqnarray}
p^{\mu}_1 - p^{\mu}_3 &=& k^{\mu}_1 \simeq z_1 p^{\mu}_1 - 
\frac{\mathbf{k}^2_1}{(1 - z_1)s} \, p^{\mu}_2 \;
+\; \mathbf{k}^{\mu}_1 \; , \nonumber \\
k^{\mu}_2 &=& \bar{z}_2 p^{\mu}_2 - \frac{\mathbf{k}^2_2} {(1-\bar{z}_2)s} \, p^{\mu}_1
+ \mathbf{k}^{\mu}_2 \; , 
\end{eqnarray}
we want to compute the $q_A q_B\rightarrow q_A q_B g$ squared matrix element in the limit of large
subenergy $s_2$
\begin{equation}
s_{2} = z_{1}s \gg s_{1} = \frac{(\mathbf{q} - z_{1} \mathbf{k}_{2})^{2}}
{z_{1} (1-z_{1})} \, , \; (\mathbf{q} = \mathbf{k}_{1} +\mathbf{k}_{2})
\, ,
\end{equation}
and fixed subenergy $s_{1}$, so as to cover the parton $A$ fragmentation
region. Since $s_{2}$ is large, we can use $\mathbf{k}$-factorization \cite{4}
to write
\begin{equation}
|\overline{M_{qg^{\ast} \rightarrow qg}}|^{2} \, = \, 
\overline{\sum}_{\textrm{spins colours}} \,
|\overline{u} (p_{3}) A^{ab}_{\mu\lambda} u(p_{1}) \varepsilon^{\mu} 2p^{\lambda}_{2}|^{2} \,
\frac{g^{6}_{s}C_{F}}{(\mathbf{k}^{2}_{2})^{2}}
\end{equation}
where, by current conservation, we can use \cite{4} the Feynman gauge
\begin{eqnarray}
A^{ab}_{\mu\lambda} &=& t^{a}t^{b} 
\frac{\gamma_{\mu} (\not{p}_{3} + \not{q})\gamma_{\lambda}}{2p_{3}q} \, 
- t^{b}t^{a} 
\frac{\gamma_{\lambda} (\not{p}_{1} - \not{q})\gamma_{\mu}}{2p_{1}q} \, + \nonumber \\
& & [t_{a},t_{b}] \frac{1}{|k^{2}_{1}|} ( (k_{1}-k_{2})_{\mu}
\gamma_{\lambda} +
(\not{k}_{2}+\not{q}) g_{\mu\lambda} -
(q+k_{1})_{\lambda} \gamma_{\mu}) \, . 
\end{eqnarray}

By using physical polarizations with $\epsilon \cdot q = \epsilon \cdot
p_{2} =0$ and the high-energy kinematics we can write the colour decomposition
\begin{equation}
A^{ab}_{\mu\lambda} \epsilon^{\mu} 2 p_{2}^{\lambda} =
A^{ab} = \{t^{a},t^{b}\} A^{(+)} + [t^{a},t^{b}]A^{(-)} \; ,
\end{equation}
with
\begin{eqnarray}
A^{\pm} &=& C^{\pm}_{1} (\not{p}_{2}p_{1} \cdot \epsilon - \not{\epsilon}
\frac{1}{2} \not{p}_{2}\not{q}) \, + \, C^{(\pm)}_{3}
(\not{p}_{2}p_{3} \cdot \epsilon +
\frac{1}{2} \not{\epsilon}\not{q}\not{p}_{2}) \; , \nonumber \\
C^{-}_{1} &=& \frac{4}{|k^{2}_{1}|} \, + \, \frac{1}{p_{1}q} \, , \;
C^{-}_{3} = \frac{1}{p_{3}q} \, - \, \frac{4}{|k^{2}_{1}|} \, , \\
C^{+}_{1} &=&  - \, \frac{1}{p_{1}q} \, , \;
C^{+}_{3} = \frac{1}{p_{3}q} \, . \nonumber
\end{eqnarray}

By performing the straightforward gamma function algebra and the
polarization sum in $D=4+2\epsilon$ dimensions, we obtain, for each
colour structure, the result
\begin{equation}
\overline{|A|^{2}} \, = \, (2s)^{2} P_{q}(\epsilon,z_{1}) \,
\left(C_{3}^{2} 2p_{3}q + C^{2}_{1} (1-z_{1})2p_{1}q +
C_{1}C_{3} (2p_{3}q (1-z_{1}) +2p_{1}q - z_{1}\mathbf{k}^{2}_{2})\right)
\end{equation}
where the splitting function of $q\rightarrow g$ type
\begin{equation}
\frac{1}{2C_{F}} P_{gq} \, = \, P_{q}(\epsilon , z) \, = \,
\frac{1}{2z} \, (1+(1-z)^{2} + \epsilon z^{2})
\end{equation}
appears to \textit{factor out} in front of the transverse momentum
dependence. Finally, by introducing phase space and coupling constant
\begin{equation}
d\phi \, =\, \pi\left(\frac{1}{(4\pi)^{2+\epsilon}}\right)^{2} \,
\frac{dz_{1} d[\mathbf{k}_{1}] d[\mathbf{k}_{2}]}
{s^{2}z_{1}(1-z_{1})} \, ,
\alpha_{s} = \frac{g^{2}_{s} \Gamma(1- \epsilon)}
{(4\pi)^{1+\epsilon}} \; ,
\end{equation}
and performing the colour algebra, we obtain
\begin{eqnarray}
\left. \frac{d\sigma^{(1)}}{d[\mathbf{k}_{1}]d[\mathbf{k}_{2}]dz_{1}} \right)_{\textrm{quark}}
 &=&
\pi \frac{(2\alpha_{s}C_{F}N_{\epsilon})^{2}}
{\mathbf{k}^{2}_{1}\mathbf{k}^{2}_{2}} \,
\frac{P_{q}(\epsilon, z_{1})}{\Gamma(1-\epsilon)} \cdot
\left[ \frac{N_{C}\alpha_{s}}{\pi} \, 
\frac{(1-z_{1})\mathbf{q} \cdot (\mathbf{q}-z_{1}\mathbf{k}_{2})}
{\mathbf{q}^{2}(\mathbf{q}-z_{1}  \mathbf{k}_{2})^{2}} 
\right. \, + \nonumber \\
&+&
\left. \frac{C_{F}\alpha_{s}}{\pi} \, \frac{z^{2}_{1}\mathbf{k}^{2}_{1}}
{\mathbf{q}^{2}(\mathbf{q}-z_{1}\mathbf{k}_{2})^{2}} \right] \, , \,
\left( N_{\epsilon} \equiv \frac{(4\pi)^{\epsilon /2}}
{\Gamma (1-\epsilon)} \right) \, .
\label{(13)}
\end{eqnarray}
This result agrees with the one extracted from Ref.(15) in the
high-energy limit.

The cross section in Eq. (13) contains two colour factors, which have a
simple interpretation, depending on the collinear singularities involved.
The $C^2_F$ term, with singularities at $\mathbf{q}^2 = 0$ $ ((\mathbf{q}
-z_{1}\mathbf{k}_{2})^2 = 0)$, comes from the Sudakov jet region, 
in which the emitted
gluon is collinear to the incoming (outgoing) quark.

On the other hand, the $C_{F}C_{A}$ term \textit{is not really singular}
 at either $\mathbf{q}^2 = 0$ or 
$(\mathbf{q}
-z_{1}\mathbf{k}_{2})^2 = 0$, except for $z_{1} = 0$, which corresponds to the
central region. It comes from the ``coherent" region in which the gluon
is emitted at angles which are large with respect to the $q_{A}q_{A^{\prime}}$
scattering angle, and is thus sensitive to the total $q_{A}q_{A^{\prime}}$ charge
$C_A$.
This is the region we are interested in, which is relevant for the energy scale,
because it tells us how the leading matrix element, valid in the central region,
is cut-off in the fragmentation region.

Since the $C_{F}C_{A}$ term in Eq.~(13) acquires the $1/\mathbf{q}^{2}$ singularity
for $|\mathbf{q}|(1-z_{1})\gg |\mathbf{k}_{1}|z_{1}$, we can write roughly,
in the quark case,
\begin{eqnarray}
\left. \frac{d\sigma^{(1)}}{d[\mathbf{k}_{1}]d[\mathbf{k}_{2}]dz_{1}} 
\right)_{\textrm{coherent}}
 &=&
\pi h^{(0)}_{q}(\mathbf{k}_{1}) h^{(0)}_{q}(\mathbf{k}_{2})
\frac{\overline{\alpha}_{s}}
{\mathbf{q}^{2}} \,
P_{q}(\epsilon, z_{1}) \Theta \left(q(1-z_{1})-k_{1}z_{1}\right) \, , \\
h^{(0)}_{q}(\mathbf{k}) &\equiv& 2 \alpha_{s} C_{F}N_{\epsilon}/\mathbf{k}^{2} \; ,
\overline{\alpha}_{s} \equiv \frac{N_{C}\alpha_{s}}{\pi} \, . \nonumber
\label{(14)}
\end{eqnarray}
The latter is the coherence effect prescription of Ref.~(14), in which
angular ordering
\begin{equation}
\frac{q}{z_{1}} \, > \, \frac{k_{1}}{1-z_{1}} \, , \; 
\theta_{q} \, > \, \theta_{p^{\prime}_{1}}
\label{(15)}
\end{equation}
replaces the smooth behaviour of the differential cross-section
in Eq.~(13). The expression (14) is actually exact upon azimuthal
averaging in $\mathbf{k}_{1}$ of Eq.~(13), at fixed $|\mathbf{q}|$.

Let me stress the point that in Eq.~(14) the $q\rightarrow g$ splitting
function $\sim P_{q} (\epsilon, z_{1})$ factors out in front
of the leading matrix element for \textit{any value of $\mathit{z_{1}}$},
not only for $z_{1} \ll 1$ as assumed originally \cite{14}. Therefore,
Eqs~(13) and (14) form the basis for the interpolation between Regge
region $(z_{1} \ll 1$, any $\mathbf{k}^{2}_{1}/\mathbf{q}^{2})$ and 
collinear region $(\mathbf{k}^{2}_{1}/\mathbf{q}^{2} \ll 1$,
any $z_{1}$), in small-$x$ equations of CCFM type \cite{14,17}.

We are now in a position to start checking Eqs~(1) and (2)
and determining scale and impact factors. Let me start with
a qualitative argument based on Eq.~(14). By introducing
the gluon (quark) rapidity $y=\log(z_{1}\sqrt{s}/q)
(Y_{1}=\log(\sqrt{s}/k_{1}))$, the restriction (15)
on half the phase space becomes
\begin{equation}
0< y < Y_{1} - \log \frac{q}{k_{1}} \Theta_{qk_{1}} \,
= \, \log (\sqrt{s}/\textrm{Max} (q,k_{1})) \, .
\label{(16)}
\end{equation}

In other words, the pure phase space $q/\sqrt{s} < z_{1}
<1$, (or $0<y<\log(\sqrt{s}/q)$) is cut-off by coherence
if $q<k_{1}$, yielding $q/\sqrt{s} < z_{1}<q/k_{1}$.

This means that, by adding the remaining phase space
related to the fragmentation region of $B$, the differential
cross section is roughly
\begin{equation}
\left. \frac{d\sigma^{(1)}}{d[\mathbf{k}_{1}]d[\mathbf{k}_{2}]} 
\right)_{\textrm{coherent}} \; \simeq \;
\pi h^{(0)}(\mathbf{k}_{1}) h^{(0)}(\mathbf{k}_{2}) \,
\frac{\overline{\alpha}_{s}}{\mathbf{q}^{2}} \;
\left( \log \frac{\sqrt{s}}{\textrm{Max} (q,k_{1})}
\, + \, \log \frac{\sqrt{s}}{\textrm{Max} (q,k_{2})}
\right)\; .
\label{(17)}
\end{equation}
Therefore, the physical scale for the energy is Max $ (q,k_{1}) \cdot$
Max $ (q,k_{2})$,
which can be roughly replaced by the expression
\begin{equation}
s_{0} (\mathbf{k}_{1},\mathbf{k}_{2}) \, = \, \textrm{Max}
(\mathbf{k}^{2}_{1}, \mathbf{k}^{2}_{2}) \; ,
\label{(18)}
\end{equation}
which has the same behaviour in the regions $q \ll k_{1} \simeq
k_{2}, q \simeq k_{2} \gg k_{1}$ and $q \simeq k_{1} \gg k_{2}$.

We have thus understood the mechanism by which the 
\textit{physical scale} is provided by the largest virtuality, in 
qualitative agreement with what we know about deep inelastic
scattering.

The coherence argument above can be made more precise
by using the exact squared matrix element in Eq.~(13), in
the phase space region connected with the fragmentation of 
$A$. In fact it is easy to check the representation
\begin{equation}
\int^{1}_{q/\sqrt{s}} dz_{1} 
\left. \frac{d\sigma}{d[\mathbf{k}_{1}]d[\mathbf{k}_{2}]dz_{1}} 
\right)_{C_{F}C_{A}}  =  \pi
\frac{h^{(0)}_{q}  (\mathbf{k}_{1}) h^{(0)}_{q}  (\mathbf{k}_{2})}
{\Gamma (1-\epsilon)}  
\frac{\overline{\alpha}_{s}}{\mathbf{q}^{2}} 
\left( \log \frac{\sqrt{s}}{\textrm{Max} (q,k_{1})}
 +  h_{q} (\mathbf{q}, \mathbf{k}_{1}) \right) \, ,
\label{(19)}
\end{equation}
where the appropriate scale has been subtracted, and
\begin{equation}
h_{q}(\mathbf{q},\mathbf{k}_{1}) \, = \, \int^{1}_{0} \,
dz_{1} \left[ P_{q} (\epsilon , z_{1})\,
\frac{(1-z_{1}) \mathbf{q} \cdot (\mathbf{q}-z_{1} \mathbf{k}_{2})}
{(\mathbf{q} - z_{1}\mathbf{k}_{2})^{2}} \, - \,
\frac{1}{z_{1}} \; + \, \log \frac{k_{1}}{q} \;
\Theta_{k_{1}q} \right]
\label{(20)}
\end{equation}
is a finite constant which, by Eq.~(14), \textit{vanishes in the $q \rightarrow 0$
limit}, thus eliminating the $q=0$ singularity in front.

The result (19), upon symmetrization, coincides with Eq.~(17) for the 
logarithmic piece. Furthermore, the constant piece $h_{q}$ can be
integrated over $\mathbf{k}_{1}$ at fixed $\mathbf{k}_{2}$ and 
interpreted as one-loop contribution to the quark $A$ impact factor
in Eq.~(2):
\begin{equation}
\left. \frac{h^{(1)}_{q} (\mathbf{k}_{2})}{h^{(0)}_{q}(\mathbf{k}_{2})}
\right)_{\textrm{real}} \, = \, \frac{\overline{\alpha}_{s}}
{\Gamma (1-\epsilon)} \, \int \, \frac{d[\mathbf{k}_{1}]}
{\mathbf{k}_{1}^{2}\mathbf{q}^{2}} \, h_{q}
(\mathbf{q}, \mathbf{k}_{1})  = 
\overline{\alpha}_{s} (\mathbf{k}^{2}_{2})^{\epsilon} \,
\left( -\frac{3}{4\epsilon} - 2\psi^{\prime} (1)-\frac{1}{4}
\right) \, .
\label{(21)}
\end{equation}

The $1/\epsilon$ pole in Eq.~(21) comes as no surprise, because
it corresponds to the collinear divergence at $k^{2}_{1}=0$ due to
the $q\rightarrow g$ transition of the initial massless quark
\begin{equation}
(2C_{F})^{-1} \hat{\gamma}_{gq}(\omega=0) \, = \,
\int^{1}_{0} \, dz \left(P_{q}(\epsilon,z) -
\frac{1}{z} \right) \, = \, - \frac{3}{4} +
\frac{1}{4} \epsilon \; ,
\label{(22)}
\end{equation}
the $\frac{1}{z}$ term being included in the leading
kernel $K_{0}$.

Eq.~(19) yields, upon symmetrization, the scale Max $ (q,k_{1}) \cdot$
Max $ (q,k_{2})$ mentioned before. The latter can be translated
into the scale Max $(k_{1}^{2},k_{2}^{2})$ of Eq.~(18) by the
change in the constant piece
\begin{equation}
\delta h_{q} (q,k_{1}) \, = \, \log \frac{k_{2}}{k_{1}} \Theta_{k_{2}k_{1}}
\, - \, \log \frac{q}{k_{1}} \Theta_{qk_{1}} \, ,
\label{(23)}
\end{equation}
which vanishes at $q=0$ also and, upon $\mathbf{k}_{1}$ integration,
corresponds to the change in impact factor
\begin{equation}
\delta h_{q}^{(1)}/h_{q}^{(0)} \, = \, \overline{\alpha}_{s}\,
\frac{1}{2} \, \psi^{\prime} (1) \, .
\label{(24)}
\end{equation}

Eqs~(18), (19), (21) and (24) summarize the one-loop results
for scale and quark impact factor at real emission level.
The virtual corrections are known \cite{5} for both $C_{F}^{2}$
and $C_{F}C_{A}$ colour factors. The Sudakov $C_{F}^{2}$ term
turns out to cancel completely with the corresponding $z_{1}$
- integrated real emission ($\omega =0$ moment) and is thus of
little concern here.

The $C_{F}C_{A}$ term, on the other hand, yields \cite{5}
\begin{eqnarray}
\left. \frac{d\sigma}{d[\mathbf{k}_{1}]d[\mathbf{k}_{2}]} 
\right)_{\textrm{virtual}}
 &=&
\pi h^{(0)}_{q}(\mathbf{k}_{1}) h^{(0)}_{q}(\mathbf{k}_{2})
2 \omega (k^{2}_{1}) \delta^{2(1+\epsilon)} (\mathbf{q})
\pi^{1+\epsilon} \, \cdot \\
\cdot \left[ \log\frac{s}{\mathbf{k}^{2}_{1}} + \left(
\frac{11}{6} - \frac{N_{F}}{N_{C}} \right)\right.
&-& 
\left. \left( 3\psi^{\prime}(1) + \frac{85}{18} - \frac{5}{9}\,
\frac{N_{F}}{N_{C}} \right) \epsilon \right] \, ,
\omega(\mathbf{k}^{2}) = - \frac{\overline{\alpha}_{s}}
{2\epsilon} \mathbf{k}^{2\epsilon} 
\frac{\Gamma^{2}(1+\epsilon)}{\Gamma(1+2\epsilon)} \; .
\nonumber
\label{(25)}
\end{eqnarray}
Here the logarithmic term provides the gluon trajectory
renormalization, which regularizes the leading BFKL kernel,
the second term in square brackets is a beta-function
coefficient which yields the running coupling for the
overall $\alpha_{s}^{2}$ factor, and half the last term
adds up to Eq.~(21) to yield the total one-loop correction
to the quark impact factor, for the scale Max $ (q,k_{1}) \cdot$
Max $ (q,k_{2})$:
\begin{equation}
h^{(1)}_{q}(\mathbf{k})/ h^{(0)}_{q}(\mathbf{k})
= \overline{\alpha}_{s}(\mathbf{k}^{2})^{\epsilon}
\left[ \left(-\frac{3}{4\epsilon} + \frac{1}{4} \right)\, +
\, \frac{1}{2} \left(\frac{67}{18} -\psi^{\prime}(1)
- \frac{5}{9}\,
\frac{N_{F}}{N_{C}} \right) \right]\, .
\label{(26)}
\end{equation}

In the final result, Eq.~(26), we have singled out the collinear
singularity due to Eq.~(22), so that its finite part is the last
term in square brackets.

The above calculation can be repeated for initial gluons. If
$D=4$ the squared $gg\ast \rightarrow gg$ matrix element can be
extracted from the helicity vertices of Ref.~(8) to yield
\begin{equation}
\left. \frac{d\sigma^{(1)}}{d[\mathbf{k}_{1}]d[\mathbf{k}_{2}]dz_{1}} 
\right)_{\textrm{gluon}} \, = \,
\frac{\pi(2\alpha_{s}N_{C})^{2}}{\mathbf{k}^{2}_{1}\mathbf{k}^{2}_{2}}
P_{g}(z_{1}) \frac{N_{C}\alpha_{s}}{\pi} \,
\frac{(1-z_{1}) \mathbf{q} (\mathbf{q}-z_{1}\mathbf{k}_{2}) + 
z^{2}_{1}\mathbf{k}^{2}_{1}}{\mathbf{q}^{2} (\mathbf{q}-z_{1}\mathbf{k}_{2})^{2}}
\, ,
\label{27}
\end{equation}
where
\begin{equation}
P_{g}(z_{1}) \; = \; \frac{1+z^{4}+(1-z)^{4}}{2z(1-z)}
\label{28}
\end{equation}
is related to the $g \rightarrow g$ splitting function. The Sudakov
and coherent regions have in this case the same colour factor, due
to the final gluons' identity.

The main difference with the quark case comes from the non singular
part of the splitting function, which yields a different constant
piece in Eq.~(19) and thus a different real emission contribution
to the gluon impact factor. The latter has the expected collinear
divergence related to $\hat{\gamma}_{gg} (\omega = 0)$, and a finite
term, whose precise determination requires generalizing Eq.~(27)
to $4+2\epsilon$ dimensions, and adding the $gg\ast \rightarrow
q\overline{q}$ contribution. The scale determination remains,
of course, unchanged.

I thus conclude that the NL one-loop calculation determines the
partonic impact factors with consistent collinear behaviour, 
depending on the choice of scale $s_{0}=$ Max $(\mathbf{k}^{2}_{1},
\mathbf{k}^{2}_{2})$. With this choice the LL logarithmic term
has no spurious $q=0$ singularities (as pure phase space would
imply) and the NL constant piece has no $q=0$ singularity at all.
Futhermore, no spurious $\mathbf{k}^{2}_{1}=0$ singularities are
present (as, for instance, the scale $s_{0}=k_{1}k_{2}$
would imply).

This means that the left-over ambiguity comes from a change
of scale keeping the behaviour of Eq.~(18) in all the regions
$k_{1} \ll k_{2}, k_{2} \ll k_{1}, q \ll k_{1} \simeq k_{2}$
as, for instance, $s_{0}=$ Max $(k_{1},q)\cdot$ 
Max $(k_{2},q)$ does. In this case the change of scale can be
reabsorbed in a change of impact factor without spoiling the 
collinear properties, as already done in Eq.~(23) for the example
just mentioned.

Once a self-consistent scale is chosen and the corresponding
impact factors are found at one-loop level, then the
two-loop calculation will determine the NL kernel unambiguously.
The corresponding results and physical implications will be
reported elsewhere.

\vspace{1cm}

I wish to thank Gianni Camici for a number of discussions and
suggestions, Keith Ellis for quite helpful correspondence on his
work, and Victor Fadin, Giuseppe Marchesini and Gavin Salam
for stimulating conversations. This work is supported in part
by M.U.R.S.T. (Italy).

\end{document}